\documentstyle[preprint,aps]{revtex}

\begin{document}      

\hoffset -1cm

\draft
\preprint{TPR-96-13, \ 
DUKE-TH-96-119, \ 
HZPP-96-6}

\title{Source Function Determined from HBT Correlations\\
       by the Maximum Entropy Principle} 

\author{Wu Yuanfang\cite{home}}

\address{Institut f\"ur Theoretische Physik, Universit\"at Regensburg,\\ 
         D-93040 Regensburg, Germany } 

\author{Ulrich Heinz\cite{home1}}

\address{Physics Department, Duke University, Durham, NC 27708-0305, USA}

\date{\today}

\maketitle 

\begin{abstract}

We study the reconstruction of the source function in space-time
directly from the measured HBT correlation function using the
Maximum Entropy Principle. We find that the problem is ill-defined
without at least one additional theoretical constraint as input. 
Using the requirement of a finite source lifetime for the
latter we find a new Gaussian parametrization of the source function
directly in terms of the measured HBT radius parameters and its
lifetime, where the latter is a free parameter which is not directly 
measurable by HBT. We discuss the implications of our results for the
remaining freedom in building source models consistent with a given
set of measured HBT radius parameters.

\end{abstract}


Relativistic heavy ion collisions are used to create small and 
shortlived regions of hot and dense hadronic matter under conditions 
resembling those of the Early Universe near the confinement phase 
transition. These little ``fireballs" expand rapidly into the vacuum, 
finally decaying into a multitude of particles whose momentum 
distributions can be measured experimentally. The only known practical 
way to extract direct information about the {\em space-time} distribution 
of the source from such {\em momentum} measurements  
is via Hanbury-Brown/Twiss (HBT) intensity interferometry which 
exploits the quantum statistical correlations in the 2-particle 
coincidence spectra. The latter are reflected in the well-known 
relationship between the momentum correlation function $C(q,K)$ 
of two identical particles and the source distribution function
$S(x,K)$ \cite{S73,P84,PCZ90,CH94}:
 \begin{equation}
 \label{corrapp}
    C({\bf q}, {\bf K}) \simeq 
    1 \pm {\vert \int d^4x\, S(x,K)\, e^{iq{\cdot}x} \vert^2 \over
         \vert \int d^4x\, S(x,K) \vert^2} \, .
 \end{equation}
Here, ${\bf K} = ({\bf p}_1 + {\bf p}_2)/2$, 
$K^0 = (E_1+E_2)/2 \approx E_K=\sqrt{m^2+{\bf K}^2}$, ${\bf q} = 
{\bf p}_1 - {\bf p}_2$, $q^0 = E_1 - E_2$. Our interest focusses
on the source function $S(x,K)$ which describes the probability that 
a particle with 4-momentum $p$ is emitted from the source at spacetime 
point $x$. It determines the effective spatial and temporal size of 
the region emitting particles with momentum $K$. The two-particle HBT
correlation function $C(q,K)$ is usually well parametrized by a 
Gaussian function in $q_\mu{=}(q_0,q_s,q_o,q_l)$ \cite{CNH95} where 
$q_0$, $q_l$, $q_o$, $q_s$ denote 
the components in time direction, parallel to the beam 
(``longitudinal" or $z$-direction), parallel to the transverse 
components ${\bf K}_\perp$ of ${\bf K}$ (``out" or $x$-direction), and 
in the remaining third cartesian direction (``side" or $y$-direction), 
respectively. For azimuthally symmetric systems it can be written 
in the ``standard form'' \cite{CSH95a} as
 \begin{equation}
 \label{gauss}
    C({\bf q}, {\bf K}) \simeq 
    1 \pm \exp\left[ -q_s^2 R_s^2({\bf K}) -q_o^2 R_o^2({\bf K})
                     -q_l^2 R_l^2({\bf K}) -2q_l q_o R_{lo}^2({\bf K})
            \right] \, ,
 \end{equation}
or in the so-called ``Yano-Koonin-Podgoretskii (YKP)" form 
\cite{YK78,P83,HTWW96,CNH95} as
 \begin{equation}
 \label{GYK}
    C({\bf q}, {\bf K}) \simeq 
    1 \pm \exp\left[ -q_\perp^2 R_\perp^2({\bf K}) 
                     -(q_l^2-q_0^2) R_\parallel^2({\bf K})
                     -(q{\cdot}U({\bf K}))^2                                                              (R_0^2({\bf K}) + R_\parallel^2({\bf K}))
            \right] \, ,
 \end{equation}
where 
 \begin{equation}
 \label{vK}
    U({\bf K}) = \gamma({\bf K}) \Bigl(1,\, 0,\, 0,\, v({\bf K}) \Bigr)
    \, , \qquad \gamma = 1/\sqrt{1-v^2}\, ,
 \end{equation}
and $q_\perp{=}\sqrt{q_s^2 + q_o^2}$. Please note that since the 
momenta $p_{1,2}$ are on-shell, the four components $q_\mu$ are not 
independent: they satisfy $K \cdot q = 0$ or 
 \begin{equation}
 \label{massshell}
  q_0 = {\beta} \cdot {\bf q} 
  \qquad {\rm with} \qquad 
  {\beta}= {\bf K}/K^0 \approx {\bf K}/E_K \, .
 \end{equation} 
${\beta}$ is the velocity of the particle pair. The mass-shell 
constraint (\ref{massshell}) is the reason
that, for azimuthally symmetric systems and in the Gaussian 
approximation \cite{CNH95}, the correlation function is completely  
determined by only four $K$-dependent functions, the HBT radius parameters
$R^2_i({\bf K})$ ($\ i=o,s,l,ol$) in the standard parametrizations,
or $R^2_i({\bf K}$) ($\ i=\perp,\parallel,0$) together with $v(K)$ in the YKP 
parametrization. As discussed in Refs. \cite{CNH95,CSH95a,HTWW96} these
measurable parameters are linear combinations of the 7 independent 
second order space-time variances (rms widths) of the source function 
$S(x,K)$. Thus the HBT correlator provides insufficient information
to reconstruct the spatial and temporal structure of the source, even in 
the Gaussian approximation. A separation of the spatial and temporal 
structure of the source is thus only possible with the help of specific
model functions for the source. So far the range of possible or even 
reasonable models is poorly understood, and its exploration 
remains an important problem in the context of HBT interferometry for 
the practitioner.  

In this note we want to try another approach, i.e. to construct a
``most likely'' or ``least biased'' class of source functions
starting directly from the (insufficient) experimental information 
in terms of the HBT radius parameters, adding a minimal amount of 
theoretical prejudice in the form of additional constraints.
It is well known that the Maximum Entropy Principle (MEP) \cite{mem}
is a powerful method in solving this kind of problem.
We will study how the additional theoretical constraints affect the
shape of this ``most likely'' source function, and to what extent such
an approach can be used with advantage in the interpretation of HBT
correlation data.

As is well known, for a given normalized probability distribution $p(x)$, 
the information entropy is defined as \cite{mem}
 \begin{equation}
 \label{infent}
    \sigma = - \int p(x)\, \ln p(x)\, d^4x \, .
 \end{equation}
In our situation the role of $p(x)$ is taken by the source function 
$S(x,K)$, where $K$ is to be treated as an additional parameter. We 
are interested in the ``most likely" spacetime distribution 
$S_K(x)\equiv S(x,K)$ for each value of $K$. Correspondingly, 
Eq.~(\ref{infent}) must be generalized as
 \begin{equation}
 \label{infentK}
    \sigma_K = - \int S(x,K)\, \ln S(x,K)\, d^4x \, .
 \end{equation}
Let us assume that we have $N$ experimentally measurable constraints:
 \begin{equation}
 \label{constraint}
  \langle g_i(x) \rangle =G_i({\bf K})\, ,
  \qquad i=1,\dots,N,
 \end{equation}
where the expectation value is to be taken with the source function:
 \begin{equation}
 \label{expval}
   \langle g_i(x) \rangle = {\int d^4x\, g_i(x)\, S(x,K) \over
                             \int d^4x\, S(x,K)} \, .
 \end{equation}
In the case of HBT interferometry the functions $g_i(x)$ are certain 
linear combinations of bilinears in the components of the coordinate 
$x_\mu$. The ``most likely" distribution function $S(x,K)$,
which has least prejudice or bias, is
the one which maximizes the entropy $\sigma_K$ under the 
constraints (\ref{constraint}). Using the method of Lagrangian multipliers 
one obtains
 \begin{equation}
 \label{SxK}
   S(x,K) = \exp\left[ - \lambda({\bf K}) 
          - \sum_{i=1}^N \lambda_i({\bf K}) \, g_i(x) \right]\, ,
 \end{equation}
where $\lambda({\bf K})$ ensures the normalization of $S(x,K)$:
 \begin{equation}
 \label{norm}
   \int d^4x\, S(x,K) = 1 \quad \Longrightarrow \quad
   e^{\lambda} = \int d^4x\, e^{-\sum_i \lambda_i\, g_i(x)}\, .
 \end{equation}
(Usually the source function is normalized to the single particle 
spectrum, $\int d^4x\, S(x,K) = E\, dN/d^3K$, but the resulting 
modifications are trivial.)
Eq.~(\ref{norm}) expresses $\lambda({\bf K})$ as a function of the other 
Lagrangian multipliers $\lambda_i({\bf K})$. These in turn are 
determined by inserting the actual values of the constraints into 
 \begin{equation}
 \label{Lag}
   -{\partial\lambda\over \partial \lambda_i} = 
  \langle g_i(x) \rangle = G_i({\bf K})\, ,
  \qquad i=1,\dots,N,
 \end{equation}
which follows directly from Eqs.~(\ref{constraint}),
(\ref{expval})
and (\ref{SxK}).

It is instructive to first consider a simple though unrealistic case. 
If the HBT correlator would simply measure the relevant geometrical 
``radii'' of the system, i.e. if
 \begin{eqnarray}
 \label{example}
  && G_1(K)=\langle \tilde x^2 \rangle \, ,
 \nonumber\\
  && G_2(K)=\langle \tilde y^2 \rangle \, ,
 \nonumber \\
  && G_3(K)=\langle \tilde z^2 \rangle \, ,
 \nonumber \\
  && G_4(K)=\langle \tilde t^2 \rangle \, ,
 \end{eqnarray}
where
 \begin{equation}
 \label{tildex}
   \tilde x^\mu \equiv x^\mu - \langle x^\mu \rangle \qquad \qquad \quad
   (\mu=0,1,2,3.)
 \end{equation}
is the deviation from the (${\bf K}$-dependent) point of maximum 
emission, $\bar x^\mu({\bf K}) = \langle x^\mu \rangle$, such that 
Eqs.~(\ref{example}) define the variances of the source 
function $S(\tilde x,K)$ in the various space-time directions,
then the ``most likely'' source distribution (\ref{SxK}) simply becomes 
 \begin{eqnarray}
 \label{Sexample}
  S(\tilde x,K)= \prod_{i=0}^3{ {1 \over \sqrt{2\pi R_i^2(K)}}
  \, \exp\left[ - {\tilde x^2_i\over 2 R^2_i(K)}\right] } .
 \end{eqnarray}
This describes a static Gaussian source with $K$-dependent radii
$R_i(K)$, $i=1,2,3$, and finite $K$-dependent lifetime $R_0(K)$.
(Remember that $K$ is the momentum of the observed particle pair.)

More generally, if the experimentally measurable constraints 
Eq.~(\ref{constraint})
are generated by bilinear functions of $\tilde x$,
 \begin{equation}
 \label{bilinear}
    g_i(\tilde x) = \tilde x_\mu\, g_i^{\mu\nu}\, \tilde x_\nu \, ,
    \qquad i=1,\dots,N,
 \end{equation}
the resulting source distribution will be a general Gaussian,
 \begin{equation}
 \label{gaussian}
    S(\tilde x,K) = e^{-\lambda({\bf K})}\, 
             \exp\left[ - \sum_{i=1}^N \lambda_i({\bf K})\, 
                       \tilde x_\mu\, g^{\mu\nu}_i \, \tilde x_\nu  \right]
    \equiv e^{-\lambda({\bf K})}\, 
      \exp\left[ - {1\over 2}\tilde x_\mu \, B^{\mu\nu}({\bf K})\, 
       \tilde x_\nu \right] \, ,
 \end{equation}
which agrees in its form with the generic Gaussian saddle point 
approximation \cite{CNH95,WSH96} of $S(\tilde x,K)$ around $\bar x({\bf K})$.  
Here, $B^{\mu\nu}=2\sum_{i=1}^N \lambda_i \, g_i^{\mu\nu}$ and $e^{-\lambda}=
\sqrt{\det B}/( 4 \pi^2)$. $B$ is a symmetric $4\times 4$
matrix and has usually 10 independent parameters. For an azimuthally 
symmetric source, only seven of these are non-zero \cite{CNH95}.

Now let us turn to the situation as it is actually encountered in nature.
If the measured correlator $C({\bf q}, {\bf K})$ is fitted only to the 
standard form (\ref{gauss}) we know that the extracted HBT radii correspond 
to the following constraints on the source \cite{CSH95a,HB95}:
 \begin{eqnarray}  
 \label{standard}
   R_s^2 &=& \langle \tilde y^2\rangle \, ,
 {\nonumber} \\
   R_o^2 &=&  \langle (\tilde x-\beta_o \tilde t)^2\rangle  \, ,
 {\nonumber} \\
   R_l^2 &=& \langle (\tilde z-\beta_l \tilde t)^2\rangle \, ,
 {\nonumber} \\
   R_{lo}^2 &=& \langle (\tilde x-\beta_o \tilde t)
                        (\tilde z-\beta_l \tilde t)\rangle \, .
 \end{eqnarray} 
In this case the normalization constant is found to be
$e^\lambda=(2\pi)^{3/2} R_s \sqrt{R^2_o R^2_l - (R^2_{ol})^2} 
\int d\tilde t$ and thus diverges. Furthermore, the corresponding matrix
 \begin{equation}
 \label{Bmatrix}
 \left( B_{\mu\nu} \right) =  {1\over \Delta}
 \left(
 \begin{array}{cccc}
   R^2_l, & 0, & -R^2_{ol}, & R^2_{ol}\beta_l-R^2_l\beta_\perp\\[3mm]
   0, & \Delta / R^2_s, & 0, & 0 \\[3mm] 
   -R^2_{ol}, & 0, & R^2_o, & R^2_{ol}\beta_\perp-R^2_o\beta_l\\[3mm]
   R^2_{ol}\beta_l-R^2_l\beta_\perp, & 0, 
     & R^2_{ol}\beta_\perp-R^2_o\beta_l,
     & R^2_l\beta_\perp^2+R^2_o\beta_l^2-2R^2_{ol}\beta_\perp\beta_l
 \end{array} 
 \right) \, ,
 \end{equation}
with
 \begin{equation}
 \label{Delta}
   \Delta = R_o^2 R_l^2 - (R_{ol}^2)^2 \, ,
 \end{equation}
has a zero determinant. Modulo the diverging normalization this 
implies that the $t$, $x$ and $z$ variances of the corresponding 
source function (\ref{gaussian}) diverge. All this implies that the 
measured HBT radii (\ref{standard}) are not sufficient to constrain 
the space-time structure of the source in all 4 space-time directions. 
The reason for this is that the observed particles are on the 
mass-shell and that therefore, as mentioned before, the time-component 
$q^0$ is not an independent variable which can be used to explore 
freely the necessary fourth dimension. Thus, based on purely 
experimental information from the standard fit alone, it is unable to 
obtain a well-defined four-dimensional space-time distribution of the 
source. We have checked that one arrives at the same conclusion 
starting from the YKP fit (\ref{GYK}) instead of the standard fit 
(\ref{gauss}). This is to be expected because the two fits are 
mathematically equivalent and correspond only a different way of 
eliminating the redundant component of $q^\mu$ using 
Eq.~(\ref{massshell}). The similar problems encountered with the 
Maximum Entropy Method in both cases reflect this equivalence.  

It is clear from this discussion that the Maximum Entropy Methods 
tends to construct from the data a source with an infinite lifetime. 
This is certainly not reasonable for the sources created in heavy-ion 
collisions, since they disperse very rapidly into the surrounding 
vacuum. But even in other situations, where it may not be a priori 
clear that the source lifetime is finite, one can easily check with 
the available experimental information whether the source is 
stationary or not. For a stationary source one sees from 
Eq.~(\ref{corrapp}) that the correlator can only be different from 
unity if $q^0={\beta} \cdot {\bf q} = 0$. This means that for pairs at 
$Y=0=K_L$, there cannot be any correlations in the ``out'' direction; 
in other words $R_o$ must be infinite. A finite value for $R_o$ is 
thus not compatible with stationarity of the source. In this case, in 
particular for heavy-ion collisions, it is thus natural to require, in 
addition to the experimental constraints (\ref{constraint}), the 
finiteness of the source lifetime: 
 \begin{equation}
 \label{time}
  \langle \tilde t^2\rangle = T^2(K).
 \end{equation}
Here $T(K)$ is an unknown, but finite $K$-dependent parameter. 
We will now show that this leads to a well-defined source function.
It will be parametrized by the chosen function $T(K)$, in addition to
the measured HBT radii.

Adding Eq.~(\ref{time}) to the constraints (\ref{constraint}) and going
again through the procedure of maximizing the entropy $\sigma_K$
we now find
 \begin{equation}
 \label{SMEP}
  S(x,K)={\exp\left[ - {\tilde y^2\over 2 R^2_s} 
        - {R^2_l (\tilde x-\beta_\perp\tilde t)^2
         + R_o^2 (\tilde z-\beta_l\tilde t)^2
         - 2 R_{ol}^2 (\tilde x-\beta_\perp\tilde t)(\tilde z-\beta_l\tilde t)
           \over 2 \Delta}
        - {\tilde t^2\over 2 T^2} \right]
        \over
        4 \pi^2 T\,R_s \sqrt{\Delta} }\, ,
 \end{equation}
with $\Delta$ from Eq.~(\ref{Delta}). Note that this source function 
depends, via the definition (\ref{tildex}) of $\tilde x^\mu$, on
four arbitrary, ${\bf K}$-dependent parameters $\bar x^\mu({\bf K})$ 
which determine the position of its saddle point $\langle x^\mu
\rangle({\bf K})$. The latter cannot be determined by measuring
particle momenta \cite{HTWW96}, since an arbitrary (even $K$-dependent)
translation of the source does not influence the latter. Thus {\em any} 
source function which is able to reproduce the measured momentum spectra 
and correlations will have this remaining ambiguity.

It is interesting to study the general physical properties of the source 
function (\ref{SMEP}) in more detail. The corresponding matrix $B$ is now 
regular and given by
 \begin{equation}
 \label{Bmatrix1}
  \left( B_{\mu\nu} \right) = {1 \over \Delta}
 \left(
 \begin{array}{cccc}
      R^2_l, & 0, & -R^2_{ol}, & R^2_{ol}\beta_l-R^2_l\beta_\perp\\ [3mm]
      0, & \Delta / R^2_s, & 0, & 0 \\ [3mm]
      -R^2_{ol}, & 0, & R^2_o, & R^2_{ol}\beta_\perp-R^2_o\beta_l\\ [3mm]
      R^2_{ol}\beta_l-R^2_l\beta_\perp, & 0, 
        & R_{ol}^2\beta_\perp-R^2_o\beta_l, 
        & R^2_l\beta_\perp^2+R^2_o\beta_l^2-2R^2_{ol}\beta_\perp\beta_l
          + \Delta / T^2 
 \end{array}  
 \right) \, .
 \end{equation}
It posesses the following important properties: first, it has the 
correct symmetries in the sense that in the limit $\beta_\perp\to 0$ 
(where the $x$ and $y$ coordinates are indistinguishable and therefore 
$R_o = R_s \equiv R_\perp$, $R_{ol}=0$, and $\Delta = R_\perp^2 R_l^2$ 
\cite{CNH95}), the terms $B_{13}$ and $B_{10}$ vanish: 
 \begin{equation}
 \label{Bmatrix2}
 \lim_{\beta_\perp \to 0} \left( B_{\mu\nu} \right) =
 \left(
 \begin{array}{cccc}
    {1\over R^2_\perp},& 0, & 0, & 0 \\ 
    0, & {1\over  R^2_\perp}, & 0, & 0\\
    0, & 0, & {1 \over R^2_l}, & -{\beta_l \over R^2_l}\\
    0, & 0, & -{\beta_l \over R^2_l}, 
       & {\beta_l^2 \over R^2_l}+{1\over T^2} 
 \end{array}  
 \right)  = 
 \left(
 \begin{array}{cccc}
    {1\over \langle \tilde x^2 \rangle},& 0, & 0, & 0 \\ 
    0, & {1\over  \langle \tilde y^2 \rangle}, & 0, & 0\\
    0, & 0, & {1 \over \langle (\tilde z - \beta_l \tilde t)^2\rangle},
       & -{\beta_l \over \langle (\tilde z - \beta_l \tilde t)^2\rangle}\\
    0, & 0, & -{\beta_l \over \langle (\tilde z - \beta_l \tilde t)^2\rangle}, 
       & {\langle (\tilde z - \beta_l \tilde t)^2\rangle +
          \beta_l^2 \langle \tilde t^2 \rangle 
          \over \langle \tilde t^2 \rangle 
          \langle (\tilde z - \beta_l \tilde t)^2\rangle}
 \end{array} 
 \right) 
 \, .
 \end{equation}
Secondly, for nonzero values of $\beta_\perp$, the nondiagonal elements 
of $B$ generating $x$-$z$ and $x$-$t$ correlations do not vanish. In 
contrast to the models studied in Ref.~\cite{CNH95}, there is actually 
no reason for them to be even small. The nonvanishing components of the 
inverse matrix, $b = \left( B^{-1} \right)$, are given by
 \begin{eqnarray}
 \label{variances}
 \langle \tilde x^2\rangle = b_{11}&=& R_o^2 + \beta_\perp^2 T^2 \, ,
          \nonumber\\
 \langle \tilde y^2\rangle = b_{22}&=& R^2_s \, ,
          \nonumber\\
 \langle \tilde z^2\rangle = b_{33}&=& R_l^2 + \beta_l^2 T^2 \, ,
          \nonumber\\
 \langle \tilde t^2\rangle = b_{00}&=&  T^2 \, ,
          \nonumber\\
 \langle \tilde z\tilde t\rangle = b_{30}&=& \beta_l T^2 \, ,
          \nonumber\\
 \langle \tilde x\tilde t\rangle = b_{10}&=& \beta_\perp T^2\, ,
          \nonumber\\
 \langle \tilde x\tilde z\rangle = b_{13}&=& R_{ol}^2 
                                           + \beta_\perp \beta_l T^2 \, .
 \end{eqnarray}
Obviously, all second order variances are now finite. Also, one easily checks 
that in the limit $\beta_\perp \to 0$ the last two components, 
$b_{10}$ and $b_{13}$, vanish as required by azimuthal symmetry.

The source (\ref{SMEP}) also gives the correct YKP fit parameters. Calculating
the latter from \cite{HTWW96}
 \begin{eqnarray}
 \label{22a}
   R_\perp^2 &=& = \langle \tilde{y}^2 \rangle \, ,
  \nonumber \\ 
   v &=& {A+B\over 2C} \left( 1 - \sqrt{1 - \left({2C\over A+B}\right)^2}
                       \right) \, ,
 \nonumber\\
   R_\parallel^2 &=&  B{-}v C,
\nonumber \\
   R_0^2 &=&  A{-}v C , 
 \end{eqnarray}
with ($\xi= \tilde x+i\tilde y$)
 \begin{eqnarray}
 \label{21a}
   A &=& \left\langle \left( \tilde t  
         - {\tilde \xi\over \beta_\perp} \right)^2 \right\rangle \, ,
 \\
 \label{21b}
   B &=&  \left\langle \left( \tilde z
         - {\beta_l\over \beta_\perp} \tilde \xi \right)^2 \right\rangle 
   \, ,
 \\
 \label{21c}
   C &=& \left\langle \left( \tilde t - {\tilde \xi\over \beta_\perp} \right)
                      \left( \tilde z - {\beta_l\over \beta_\perp}
                             \tilde \xi \right) \right\rangle \, ,
 \end{eqnarray}
evaluated with the variances (\ref{variances}) calculated from the source 
(\ref{SMEP}), one easily verifies that they satisfy the relations with the
standard parameters (\ref{standard}) given in Eqs.(21,22) of 
Ref.~\cite{HTWW96}. However, since the experimentally unmeasurable 
parameter $T$ from the additional theoretical constraint (\ref{time}) is
specified in a fixed frame, while the YKP parameter $R_0$ measures the source
lifetime in the comoving Yano-Koonin ($v=0$) frame, there is no simple relation
between $R_0$ and $T$. For the same reason, the $T$-dependence of the 
space-time variances (\ref{variances}) will look less simple in any other
frame which moves with a fixed, but non-zero longitudinal velocity relative
to the frame in which (\ref{time}) is specified.

It is interesting to compare the matrix (\ref{Bmatrix1}) with the 
one extracted in Ref.~\cite{CNH95} from the constraints (\ref{standard}) 
under the additional assumption that the $x$-$z$ and $x$-$t$ 
correlations $B_{xz}$ and $B_{xt}$ are small and can be neglected:
 \begin{equation}
 \label{BmatrixCNH}
  \left( B_{\mu\nu} \right) = 
 \left(
 \begin{array}{cccc}
   {1\over 2R^2_s},& 0, & 0, & 0 \\ 
    0, & {1\over 2 R^2_s}, & 0, & 0\\
    0, & 0, & {R_o^2 - R_s^2 \over (R_o^2-R_s^2)R_l^2 - (R_{ol}^2)^2},
       & - {(R_o^2-R_s^2)\beta_l - R_{ol}^2 \beta_\perp \over 
            (R_o^2-R_s^2)R_l^2 - (R_{ol}^2)^2}\\
    0, & 0, &  - {(R_o^2-R_s^2)\beta_l - R_{ol}^2 \beta_\perp \over 
                  (R_o^2-R_s^2)R_l^2 - (R_{ol}^2)^2},
       & {R_l^2 \beta_\perp^2 + (R_o^2-R_s^2)\beta_l^2 
          - 2 R_{ol}^2 \beta_\perp \beta_l \over
          (R_o^2-R_s^2)R_l^2 - (R_{ol}^2)^2}
 \end{array} \, 
 \right) .
 \end{equation}
This has already the block-diagonal shape (\ref{Bmatrix2}), even for 
non-zero $\beta_\perp$. It is a regular matrix, with non-zero determinant:
 \begin{equation}
 \label{detB}
   \det B = {\beta_\perp^2 \over R_s^4 \left[
             (R_o^2 - R_s^2) R_l^2 - R_{ol}^4 \right]} \, .
 \end{equation}
This means that the above additional assumptions were also sufficient to 
guarantee a well-defined and finite source function in the Gaussian 
approximation, using only the measurable constraints (\ref{standard}). These 
assumptions can be justified for sources without collective transverse 
expansion, where they are satisfied exactly \cite{CNH95,WHTW96}. For
non-zero transverse flow, however, they are usually violated 
\cite{HTWW96,WHTW96}. Our solution (\ref{Bmatrix1}) does not make such 
assumptions, at the expense of introducing an additional unknown parameter
function $T(K)$. It is interesting to note that in the limit $\beta_\perp\to 0$
the matrix (\ref{BmatrixCNH}) takes a form which differs from (\ref{Bmatrix2}):
 \begin{equation}
 \label{BmatrixCNH2}
 \lim_{\beta_\perp \to 0} \left( B_{\mu\nu} \right) =
 \left(
 \begin{array}{cccc}
    {1\over \langle \tilde x^2\rangle},& 0, & 0, & 0 \\ 
    0, & {1\over  \langle \tilde y^2 \rangle}, & 0, & 0\\
    0, & 0, & {\langle \tilde t^2 \rangle \over 
               \langle \tilde t^2 \rangle \langle \tilde z^2 \rangle 
             - \langle \tilde t \tilde z \rangle^2}, 
       & - {\langle \tilde t \tilde z \rangle \over 
            \langle \tilde t^2 \rangle \langle \tilde z^2 \rangle 
          - \langle \tilde t \tilde z \rangle^2}\\
    0, & 0, &-{\langle \tilde t \tilde z \rangle \over 
               \langle \tilde t^2 \rangle \langle \tilde z^2 \rangle 
             - \langle \tilde t \tilde z \rangle^2},
       & {\langle \tilde z^2 \rangle \over 
          \langle \tilde t^2 \rangle \langle \tilde z^2 \rangle 
        - \langle \tilde t \tilde z \rangle^2}
 \end{array} 
 \right) \, .
\end{equation} (Of course, $\langle \tilde x^2 \rangle = \langle \tilde y^2
\rangle \equiv R_\perp^2$ in this limit.) The corresponding source 
(\ref{gaussian}) thus differs from the $\beta_\perp \to 0$ limit of 
(\ref{SMEP}). Still, both sources reproduce the same constraints 
(\ref{standard}). This means that along the particular space-time directions
in which the HBT correlator is able to probe the source $S(x,K)$ via the 
restricted (see Eq.~(\ref{massshell})) Fourier transform (\ref{corrapp}), both
sources have the same Gaussian curvature. 

In this note we have taken the basic philosophy of HBT interferometry,
namely to obtain information on the space-time structure of the source from
momentum measurements only, to the limit. With the help of the Maximum 
Entropy Principle we derived the source $S(x,K)$ directly from the experimental
constraints provided by the measured HBT correlator, supplemented by one
minimal but necessary additional theoretical constraint limiting the 
source lifetime. This additional constraint is required to render the problem
well-defined, because the kinematic restrictions imposed by the mass-shell
constraint (\ref{massshell}) generally prohibit the inversion of the relation
(\ref{corrapp}) between $S(x,K)$ and $C({\bf q},{\bf K})$. Our resulting source
is a Gaussian in the space-time coordinates, parametrized by the ${\bf
K}$-dependent HBT radii (\ref{standard}) and the lifetime parameter $T({\bf
K})$, as well as by the (unmeasurable) ${\bf K}$-dependent position of its
saddle point $\langle x^\mu\rangle$. It differs from the Gaussian approximation
given in Ref.~\cite{CNH95} which was derived using different additional
theoretical constraints, namely the vanishing of the $x$-$t$ and $x$-$z$ 
correlations $\langle \tilde x \tilde t \rangle$, $\langle \tilde x \tilde z
\rangle$ independent of the pair momentum. Both sources reproduce, however, the
same (measured) HBT correlation function. Our result thus provides not only a
possibly useful new source parametrization, but more importantly an explicit
and interesting example illustrating the remaining freedom in the space-time
structure of the source left open by even the most accurate HBT measurements.

This work was supported by grants from DFG, NSFC, BMBF and GSI. The 
authors are grateful to Urs Wiedemann for very constructive 
discussions during the initial part of this work and for a careful 
reading of the manuscript. U.H. would like to thank B. M\"uller and 
the Duke University Physics Department for their warm hospitality. 
Y.W. would like to thank Liu Lianshou for his valuable comments on the 
draft.

\end{document}